\documentclass[letterpaper, 10 pt, conference]{ieeeconf}
\usepackage[T1]{fontenc}
\usepackage[latin9]{inputenc}
\usepackage{float}
\usepackage{amsmath}
\usepackage{amssymb}
\usepackage{graphicx}
\usepackage{subcaption}

\def\baselinestretch{0.97}
\IEEEoverridecommandlockouts                              
\makeatletter

\providecommand{\tabularnewline}{\\}

\makeatother

\begin{document}

\title{Modelling and Fast Terminal Sliding Mode Control for Mirror-based Pointing Systems}
\author{Ansu Man Singh, Manh Duong Phung, Q. P. Ha\\
University of Technology Sydney, Australia.\\
{\tt\small \{AnsuMan.Singh, manhduong.phung,Quang.Ha\}@uts.edu.au}
}
\maketitle

\begin{abstract}
In this paper, we present a new discrete-time Fast Terminal Sliding
Mode (FTSM) controller for mirror-based pointing systems. We first derive the decoupled model of those systems and then estimate the parameters using a nonlinear least-square identification method. Based on the derived model, we
design a FTSM sliding manifold in the continuous domain. We then exploit the Euler discretization on the designed FTSM sliding surfaces to synthesize a discrete-time controller. Furthermore, we improve the transient dynamics of the sliding surface by adding a linear term. Finally, we prove the stability of the proposed controller based on the Sarpturk reaching condition. Extensive simulations, followed by comparisons with the Terminal Sliding Mode (TSM) and Model Predictive Control (MPC) have been carried out to evaluate the effectiveness of the proposed approach. A comparative study with data obtained from a real-time experiment was also conducted. The results indicate the advantage of the proposed method over the other techniques.
\end{abstract}

\section{Introduction}

Sensors like LIDAR plays an important role in robot navigation and
mapping. However, one of the bottlenecks in the advancement of the
navigational technology comes from the slower responses of such sensors.
Particularly, when the sensors undergo rotation, their speeds are limited
because of their inertia. One of the techniques employed to solve
the issue is using a light-weight mirror directly above the sensor,
which undergoes rotation to provide the sensing capability \cite{wood2012novel}.
As a result of its low inertia, sensing speed can be improved.

Mirror-based pointing sensors can be integrated with other sensors
such as thermal camera, and provide a wide range of applications. For
instance, in \cite{okumura2011high}, vision cameras are used with
such systems to track the motion of table-tennis balls.
 
An important component of such a pointing system is the control system. Various control techniques can be implemented and studied in such systems. However, proportional, integral, and derivative (PID) controllers are mostly implemented in such systems because of their simplicity in design and implementation. For instance, in \cite{wood2012novel} PID controllers are implemented for motion control of the mirror.\par

The mirror-based pointing sensors are, however, not free from noises, disturbances, and uncertainties which arise
from various sources such as frictions, nonlinearities, unmodelled
dynamics, and so on. As a result of the disturbances, the performance
of the system is affected, particularly, in high-speed applications
mentioned previously. To address the issue, this paper presents a
robust tracking control system based on discrete-time Fast Terminal
Sliding Mode Control (FTSM).

For the implementation of the proposed controller, first, a decoupled
dynamics for the system is developed, followed by, the identification
of its parameters. Then, sliding surfaces based on FTSM manifold are
designed. Since the control system is implemented in digital hardware,
the sliding manifolds are discretized using Euler's discretization.
Then, a discrete-time control law is synthesized. Finally, proof of the stability
of the control system is also provided by utilizing Sarpturk reaching
condition.

The organization of the paper is as follows. The construction
of the mirror-based pointing sensor and its dynamics modeling are presented
in Section \ref{sec:Mirror-based-pointing-system},
followed by the control system design in Section \ref{secControl-system-Design}.
Simulation results of the proposed method and its comparison with other methods are presented in Section \ref{sec:Simulation-results}. Finally, Section \ref{sec:Conclusion} draws the paper's conclusion and suggests its future work.

\section{Mirror-based pointing system\label{sec:Mirror-based-pointing-system} }

\begin{figure}[h]
\begin{centering}
\includegraphics[width=0.9\columnwidth]{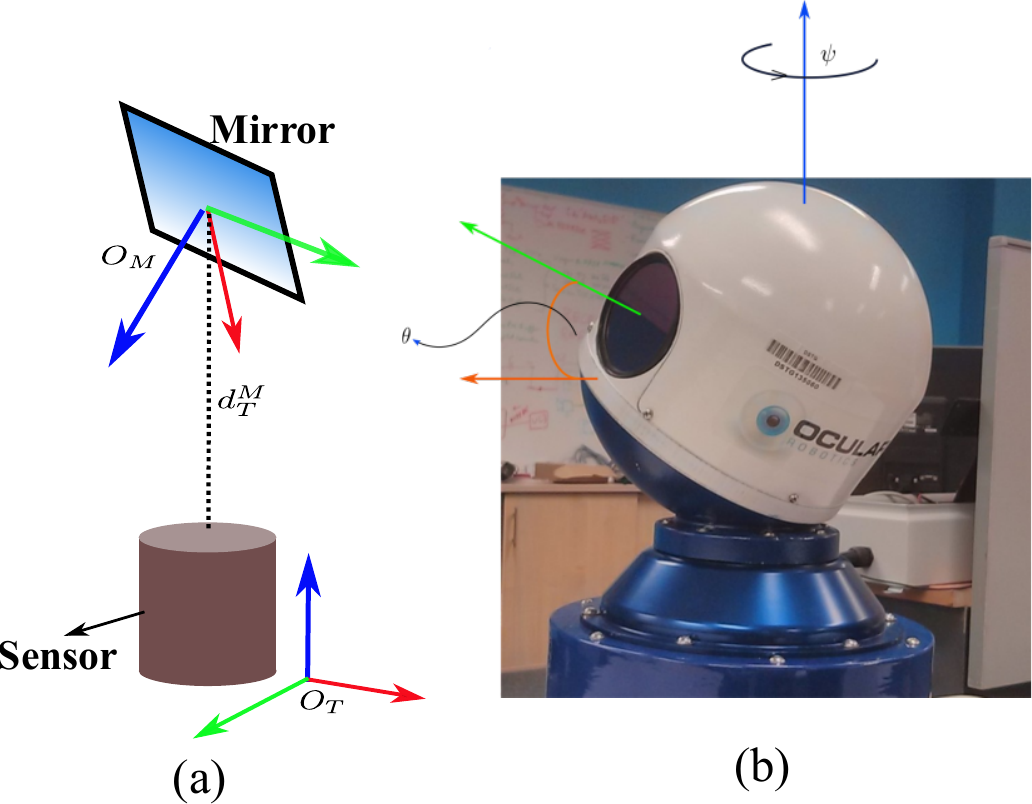}
\par\end{centering}
\caption{Mirror based sensor\label{fig:Mirror-based-sensor}}
\end{figure}
A mirror-based pointing sensor, as shown in Fig. \ref{fig:Mirror-based-sensor} (a), has a light-weight mirror which is placed directly above the sensor. The sensor can be of any type, such as thermal cameras, vision cameras, etc. The advantage of this type of pointing system is that the mirror, which has low momentum, undergoes rotational motion to provide sensing capability. As a result, the dynamic response of the sensor is tremendously improved. In this paper, we consider the pointing sensor developed and commercialized by Ocular Robotics Pty. Ltd., which is also known as RobotEye.

Owing to the faster responses of such devices, they can improve thermoelastic stress analysis (TSA) of mechanical structures for their structural health monitoring and fatigue analysis. Particularly, such pointing sensors can provide blur-free images of structure-under-test during TSA by compensating its motion \cite{Singh2018}.

This type of sensor can be represented by two variables namely azimuth
and the elevation angles, as shown in Fig. \ref{fig:Mirror-based-sensor}
(b). The azimuth angle ($\psi$) represents the rotation of the sensor head
about the vertical axis as depicted in the figure. Similarly, the
elevation angle ($\theta$) represents the angle made by the viewing direction
with respect to the horizontal plane.

\subsection{Coupled dynamics of the system}
The motion dynamics of the system can be represented by the following
MIMO transfer function, i.e.
\begin{equation}
\left[\begin{array}{c}
\theta(s)\\
\psi(s)
\end{array}\right]=\left[\begin{array}{cc}
H_{11}(s) & H_{12}(s)\\
H_{21}(s) & H_{22}(s)
\end{array}\right]\left[\begin{array}{c}
\theta_{u}(s)\\
\psi_{u}(s)
\end{array}\right],\label{eq:azimuth_elevation_H_model}
\end{equation}
where $s$ represents the laplace variable, $\theta_{u}(s)$ and $\psi_{u}(s)$
represent the inputs to the system, and $\theta(s)$ and $\psi(s)$
are the outputs of the system. Similarly, 
\[
\begin{gathered}\begin{aligned}H_{11}(s) & =\left.\frac{\theta(s)}{\theta_{u}(s)}\right|_{\psi_{u}=0},\\
H_{12}(s) & =\left.\frac{\theta(s)}{\psi_{u}(s)}\right|_{\theta_{u}=0},
\end{aligned}
\end{gathered}
\,\begin{gathered}\begin{aligned}H_{21}(s) & =\left.\frac{\psi(s)}{\theta_{u}(s)}\right|_{\psi_{u}=0},\\
H_{22}(s) & =\left.\frac{\psi(s)}{\psi_{u}(s)}\right|_{\theta_{u}=0}.
\end{aligned}
\end{gathered}
\]
It is noted that system (\ref{eq:azimuth_elevation_H_model})
represents the coupling of the azimuth and elevation dynamics via
$H_{12}$ and $H_{21}$ transfer functions. 

\subsection{Decoupling of elevation and azimuth dynamics}
From (\ref{eq:azimuth_elevation_H_model}), dynamics of elevation
angle can be represented as:
\[
\theta(s)=H_{11}(s)\theta_{u}(s)+H_{12}(s)\psi_{u}(s),
\]
or,
\begin{equation}
\theta(s)=H_{11}(s)\theta_{u}(s)+D_{\psi}(s),
\end{equation}
where $D_{\psi}(s)=H_{12}(s)\psi_{u}(s)$ represents the disturbance due to the coupling
from $\psi$. Similarly, one can also express $\psi(s)$ as 
\begin{equation}
\psi(s)=H_{22}(s)\psi_{u}(s)+D_{\theta}(s),
\end{equation}
where $D_{\theta}(s)=H_{21}(s)\theta_{u}(s)$ represents the disturbance arising from the
coupling. Here, $D_{\psi}(s)$ and $D_{\theta}(s)$ are assumed to
be bounded, i.e. $\left|D_{\psi}(s)\right|<\mu_{\psi}$ and $\left|D_{\theta}(s)\right|<\mu_{\theta}.$

\subsubsection{Identification of $H_{11}(s)$ and $H_{22}(s)$}

The $H_{11}(s)$ and $H_{22}(s)$ can be represented by a second-order
system, whose transfer function is given by
\begin{equation}
\begin{gathered}\begin{aligned}H(s) & =\frac{b_{0}}{s^{2}+a_{1}s+a_{2}}.\end{aligned}
\end{gathered}
\label{eq:second-order-transfer-fxn}
\end{equation}
In order to identify the parameters $b_{0},$ $a_{1},$ and $a_{2}$
for $H_{11}(s)$ and $H_{22}(s)$, input-output datas were collected
from the pointing sensor. Then, nonlinear
least-square methods were applied, which are provided as an application
program interface (API) in Matlab (see \cite{garnier2003continuous}
for the details on the methods). The identified parameters are listed
in Table \ref{tab:The-identified-parameters-H11-H22}.

\begin{table}[H]
\begin{centering}
\caption{The identified parameters of $H_{11}$ and $H_{22}$.\label{tab:The-identified-parameters-H11-H22}}
\begin{tabular}{c|c|c}
Parameters & $H_{11}$ & $H_{22}$\tabularnewline
\hline 
$b_{0}$ & 3581 & 3317\tabularnewline
\hline 
$a_{1}$ & 59.6 & 58.6\tabularnewline
\hline 
$a_{2}$ & 3568 & 3310\tabularnewline
\hline 
\end{tabular}
\par\end{centering}
\end{table}

\begin{figure}[h]
\begin{centering}
\includegraphics[width=0.95\columnwidth]{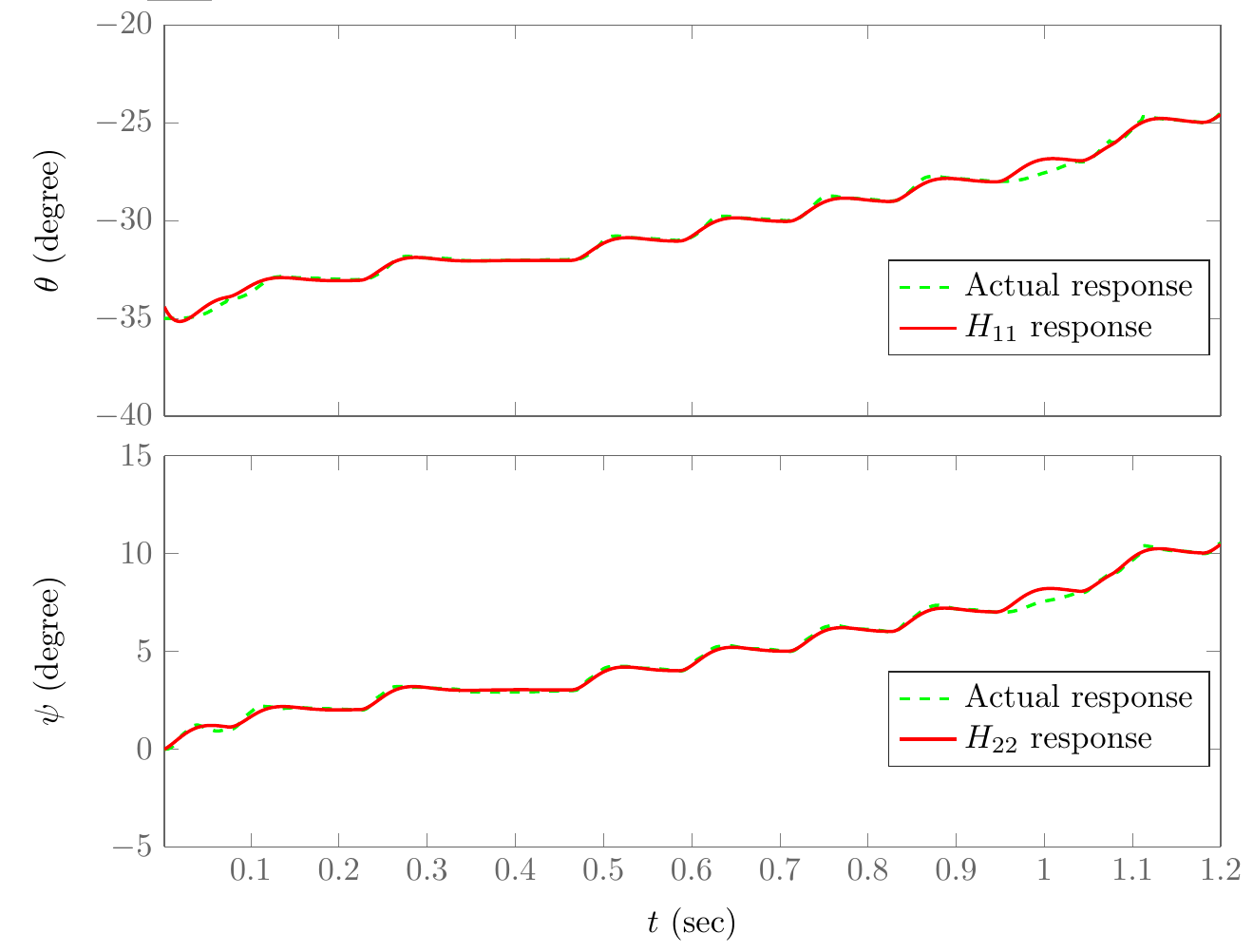}
\par\end{centering}
\caption{Responses of the identified models, i.e. $H_{11}$ and $H_{22}.$ \label{fig:Responses-of-the-identified-models} }
\end{figure}

Verification of the identified models are presented in the Fig. \ref{fig:Responses-of-the-identified-models}. The figure shows the comparison of the responses of $H_{11}$ and
$H_{22}$ with respect to the actual responses. From the figure, it
is clear that the outputs of the models are close to the actual. The
accuracy of the model response is around 93 and 94 \%, respectively,
for $H_{11}$ and $H_{22}$, which is measured in terms of normalized root mean square error (NRMSE). The performance index is defined as
\begin{equation}
\text{\,NRMSE}=\left(1-\frac{\sqrt{\frac{\left\{ \sum_{i=1}^{N_s}\left(y_{i}-\hat{y}_{i}\right)^{2}\right\} }{N_s}}}{\sqrt{\sum_{i=1}^{N_s}\left(y_{i}-\bar{y}\right)^{2}}}\right),
\end{equation}
where $y_{i\,(i=1\hdots N_s)}$ are the actual outputs, $\hat{y}_{i\,(i=1\hdots N_s)}$ are the predicted outputs of the  estimated models, $\bar{y}$ is the mean of the actual outputs, i.e. $y_{i\,(i=1\hdots N_s)}$, and $N_s$ is the number of samples.  
\subsubsection{State-space representation of $H_{11}$ and $H_{22}$:}
The second order transfer functions $H_{11}$ and $H_{22}$ can be
represented in the state-space form as
\begin{equation}
\begin{gathered}\begin{aligned}\dot{x}_{1} & =x_{2}(t)\\
\dot{x}_{2} & =-a_{1}x_{1}(t)-a_{2}x_{2}(t)+b_{0}u(t)+d(t),
\end{aligned}
\end{gathered}
\label{eq:state-space-model-of-REsystem}
\end{equation}
where $x_{1}(t)\in\mathbb{R}$ and $x_{2}(t)\in\mathbb{R}$ are the states
of the system, and $u(t)\in\mathbb{R}$ is the input to the system. The
lumped term $d(t)\in\mathbb{R}$ represents the disturbance due to the coupling and other factors, such as unmodelled dynamics of the system. In this paper, $d(t)$ is assumed
to be bounded, i.e. $\left|d(t)\right|<\mu.$

\section{Control System Design\label{secControl-system-Design}}

\subsection{Fast Terminal Sliding Mode}

In order to track a reference signal $r(t)$ for system (\ref{eq:state-space-model-of-REsystem}),
one can design Fast Terminal Sliding Mode (FTSM) based sliding functions
which are given by: 

\begin{equation}
\begin{gathered}\begin{aligned}\sigma_{1} & =x_{1}-r(t)\\
\sigma_{2} & =\dot{\sigma}_{1}+\alpha\sigma_{1}+\beta\sigma_{1}^{q_{1}/p_{1}},
\end{aligned}
\end{gathered}
\label{eq:sliding_surfaces_continuousTime}
\end{equation}
where $q_{1}>0$ and $p_{1}>0$ are odd integers such that $0<\frac{q_{1}}{p_{1}}<1,$
$\alpha>0,$ and $\beta>0.$ 

According to the theory of FTSM \cite{yu2002},
$\sigma_{1}$ has finite-time reachability property. In other words,
when $\sigma_{2}=0$, $\sigma_{1}$ reaches equilibrum in finite-time.
Now, the control law for system (\ref{eq:state-space-model-of-REsystem})
using FTSM manifold is obtained as:

\begin{equation}
\begin{gathered}\begin{aligned}u & =-\frac{u_{0}}{b_{0}}\\
u_{0} & =
-a_{1}x_{1}-a_{2}x_{2}-\ddot{r}+\alpha\dot{\sigma}_{1}+\beta\frac{q_{1}}{p_{1}}\sigma^{\frac{q_{1}-p_{1}}{p_{1}}}\dot{\sigma}_{1}\\
& \quad \; +\Phi\sigma_{2}+K\text{sign}\left(\sigma_{2}\right),
\end{aligned}
\end{gathered}
\label{eq:FTSM_control_Input}
\end{equation}
where $K>\mu$ and $\Phi>0.$

\subsection{Discrete-time FTSM control input synthesis}
There are many research papers on continuous-time FTSM, but very few studies deal with the synthesis and analysis of discrete-time FTSM. In this regard, Shihua Li et al., in \cite{li2014discrete}, provided a comprehensive study on the control system design methodology based on Terminal Sliding Mode (TSM), followed by its analysis in steady-state condition by Behera et al. in \cite{behera2015steady}. In this paper, we follow their methodology to design discrete-time FTSM controller. Here, we apply the Euler's discretization to (\ref{eq:state-space-model-of-REsystem}), 
as in \cite{galias2007euler}, which leads to:
\begin{equation}
\begin{gathered}\begin{aligned}x_{1}[n+1] & =x_{1}[n]+Tx_{2}[n]\\
x_{2}[n+1] & = a_{1}Tx_{1}[n]+\left(1+a_{2}T\right)x_{2}[n]\\
& \quad \; +b_{0}Tu[n]+Td[n],
\end{aligned}
\end{gathered}
\label{eq:discretised_state_space}
\end{equation}
where $T$ is the sampling period. Furthermore, the discretization of
the FTSM sliding surfaces  (\ref{eq:sliding_surfaces_continuousTime})
results in: 

\begin{equation}
\begin{gathered}\begin{aligned}\sigma_{1}[n] & =x_{1}[n]-r[n]\\
\sigma_{2}[n] & =\Delta\sigma_{1}[n]+\alpha\sigma_{1}[n]+\beta\sigma_{1}^{q_{1}/p_{1}}[n],
\end{aligned}
\end{gathered}
\label{eq:discretized_sliding_surface}
\end{equation}
where $\alpha>0,$ $\beta>0,$ and $q_{1}$ and $p_{1}$ are odd integers
such that $q_{1}<p_{1}.$ Here, $\Delta$ is the discrete-time approximation
of differential operator \cite{li2014discrete}, which is also known
as the forward differential operator:

\begin{equation}
\Delta\left(x[n]\right)=\frac{x[n+1]-x[n]}{T}.
\end{equation}
 By applying differential operator, $\sigma_{1}$ can also be represented
as: 
\begin{equation}
\Delta\sigma_{1}[n]=x_{2}[n]-\Delta\left(r[n]\right).\label{eq:differential_operator_s1}
\end{equation}
Similarly, $\Delta\left(\sigma_{1}^{q_{1}/p_{1}}\right)$ is given
by: 
\begin{equation}
\Delta\left(\sigma_{1}^{q_{1}/p_{1}}\right)=\frac{q_{1}}{p_{1}}\sigma_{1}^{q_{1}/p_{1}-1}\Delta\left(\sigma_{1}\right).\label{eq:derivative_terminal_exp}
\end{equation}
Now, we propose the following control law for the system (\ref{eq:discretised_state_space}):

\begin{equation} 
\begin{gathered}\begin{aligned}u & =-\frac{u_{0}}{b_{0}}\\
u_{0} & = -a_{1}x_{1}[n]-a_{2}x_{2}[n]-\Delta^{2}(r[n])\\
& \quad \; +(\alpha+\beta\frac{q_{1}}{p_{1}}\sigma_{1}^{q_{1}/p_{1}-1})\Delta\left(\sigma_{1}^{q_{1}/p_{1}}\right)\\
& \quad \; +\Phi\sigma_{2}+K\text{sign}\left(\sigma_{2}\right),
\end{aligned}
\end{gathered}
\label{eq:control_Input_FTSM_discrete}
\end{equation}
where $K>0$ and $\frac{1}{T}>\Phi>0.$ The role of $K$ in (\ref{eq:control_Input_FTSM_discrete}) is to improve robustness in the presence of noise and disturbances. Furthermore, it should be noted that for the discrete-time sliding surface $\sigma_2$ dyanamics does not goes to 0 as in continous-time, but remains within a bounded region, which is also known as quasi sliding mode band \cite{gao1995discrete}. The bounded region depends on $K$ \cite{behera2015steady}.

\begin{figure}[h]
	\begin{center}
\includegraphics[width=0.8\columnwidth]{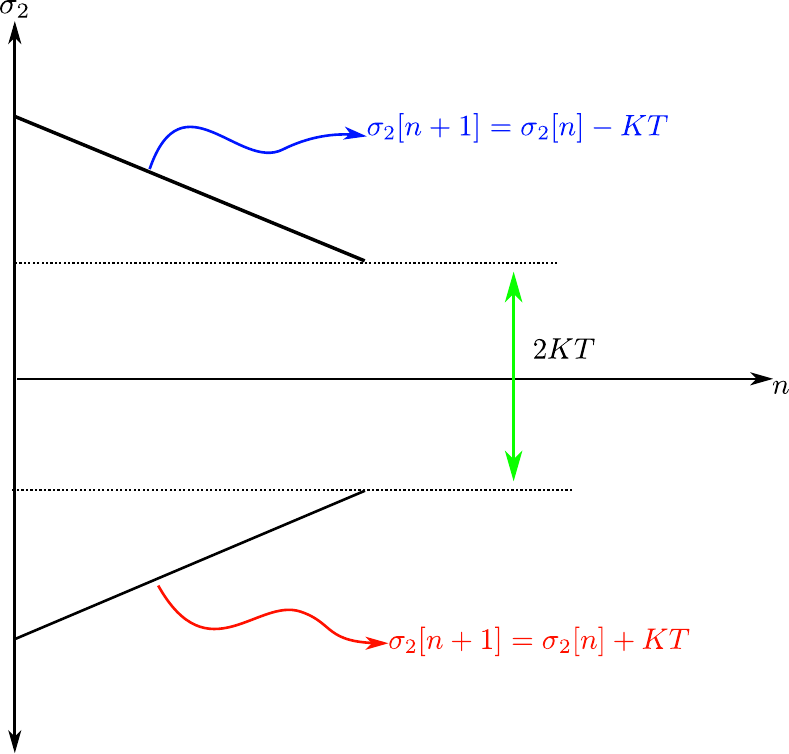}
	\end{center}
	\caption{Transient analysis of $\sigma_2$ \label{fig:transient-of-sliding-surfaces}}
	
\end{figure}

Here, it should be noted that we have introduced a new term, $\Phi \sigma_2$, to the control law (\ref{eq:control_Input_FTSM_discrete}). The rationale is to improve the transient response of the dynamics of $\sigma_2$. To understand the phenomena, consider the dynamics  of $\sigma_2$ resulting from the application of the control input without the $\Psi \sigma_2[n]$ in (\ref{eq:control_Input_FTSM_discrete}), as in \cite{li2014discrete}, i.e.
\begin{equation}
\sigma_{2}[n+1]-\sigma_{2}[n]= -KT\text{sign}\left(\sigma_{2}[n]\right).
\end{equation}
Now, consider $\sigma_{2}$ outside the boundary region or when the system is in transient, i.e. $\left|\sigma_{2}\right|>2KT$.
In this region, let us consider two cases, i.e. $\sigma_{2}>0$ and $\sigma_{2}<0$.
Expression of $\sigma_2$ for both cases are given by
\begin{equation}
\begin{cases}
\sigma_2[n+1]=\sigma_2[n]-KT, & \sigma_2>0\\
\sigma_2[n+1]=\sigma_2[n]+KT, & \sigma_2<0,
\end{cases}
\end{equation}
and the plot of $\sigma_2$ in these regions is shown in Fig. \ref{fig:transient-of-sliding-surfaces}. It
is obvious from the equations that $K$ affects the slope of $\sigma_{2}$ 
trajectory. As a result, higher values of $K$ result in faster response
of the system. However, it also means the region of oscillation in
the steady state condition is also higher. Therefore, there exists a tradeoff while selecting values for $K$, and hence it limits the transient performance of $\sigma_2$.  
\subsection*{Reaching condition for sliding surfaces}

Owing to the sampling process, the reaching condition for discrete-time
system cannot be directly mapped from its continuous-time counterpart
by using the concept of equivalent control \cite{shtessel2014sliding}.
Many studies have been conducted in the past to establish the reaching
condition. One such condition was proposed by Sarpturk et al.
\cite{sarpturk1987stability}, which is given by 
\begin{equation}
\left|\sigma[k+1]\right|<\left|\sigma[k]\right|.\label{eq:Sarpturk_condition}
\end{equation}
The condition can also be represented as
\[
\begin{gathered}\left[\sigma[k+1]-\sigma[k]\right]\text{sign}\left(\sigma[k]\right)<0\\
\left[\sigma[k+1]+\sigma[k]\right]\text{sign}\left(\sigma[k]\right)>0,
\end{gathered}
\]
which indicates that a system satisfying the condition (\ref{eq:Sarpturk_condition})
converges towards the sliding surface, and then it remains within
a bounded region after that. The bounded region is also known as quasi
sliding mode band \cite{zheng2006design}. 

\subsection{Stability Analysis }
Equation (\ref{eq:discretized_sliding_surface}) can be represented
as:
\begin{equation}
\begin{aligned}\sigma_{2}[n] & =x_{2}[n]-\Delta(r[n])+\alpha\sigma_{1}[n]+\beta\sigma_{1}^{q_{1}/p_{1}}[n].\end{aligned}
\label{eq:s2_equation}
\end{equation}
Similarly, one can get
\begin{equation}
\begin{aligned}
\sigma_{2}[n+1] & =
x_{2}[n+1]-\Delta(r[n+1])+\alpha\sigma_{1}[n+1]\\
& \quad \; +\beta\sigma_{1}^{q_{1}/p_{1}}[n+1].
\end{aligned}
\label{eq:s2n_plus1}
\end{equation}
Now, applying the forward differential operator to $\sigma_{2}[n]$,
and then to (\ref{eq:derivative_terminal_exp}) and (\ref{eq:differential_operator_s1}),
one obtains 
\begin{equation}
\begin{aligned}\frac{\sigma_{2}[n+1]-\sigma_{2}[n]}{T} & =
x_{2}[n+1]-x_{2}[n]-\Delta^{2}\left(r[n]\right)\\
& \quad \; +\left(\alpha+\beta\frac{q_{1}}{p_{1}}\sigma_{1}^{q_{1}/p_{1}-1}\right)\Delta\left(\sigma_{1}\right).
\end{aligned}
\label{eq:differences_sliding_surfaces}
\end{equation}
By applying (\ref{eq:discretised_state_space}) and the control
law (\ref{eq:FTSM_control_Input}) to (\ref{eq:differences_sliding_surfaces}), one obtains
\[
\begin{aligned}\sigma_{2}[n+1]-\sigma_{2}[n] & =-\Phi T\sigma_{2}[n]-KT\text{sign}\left(\sigma_{2}[n]\right),\end{aligned}
\]or,
\[
\begin{aligned}\sigma_{2}[n+1] & =(1-\Phi T)\sigma_{2}[n]-KT\text{sign}\left(\sigma_{2}[n]\right).\end{aligned}
\]
Now, taking the absolute values on both sides of the equation leads
to 
\[
\left|\sigma_{2}[n+1]\right|=\left|(1-\Phi T)\sigma_{2}[n]-KT\text{sign}\left(\sigma_{2}[n]\right)\right|,
\]
or,
\[
\left|\sigma_{2}[n+1]\right|=\left|(1-\Phi T)\sigma_{2}[n]\right|-\left|KT\text{sign}\left(\sigma_{2}[n]\right)\right|.
\]
Since $0<\left(1-\Phi T\right)<1$, we have$\left|(1-\Phi T)\sigma_{2}[n]\right|-\left|KT\text{sign}\left(\sigma_{2}[n]\right)\right|<\left|\sigma_{2}[n]\right|.$
As a result,
\[
\left|\sigma_{2}[n+1]\right|<\left|\sigma_{2}[n]\right|.
\]
Hence, the system satisfies the Sarptuk condition (\ref{eq:Sarpturk_condition}).
Therefore, the system converges towards sliding surfaces and remains
bounded. Similarly, once $\sigma_{2}[n]$ is bounded, by using (\ref{eq:discretized_sliding_surface}) it can be concluded that $\sigma_{1}[n]$
or the error is also bounded \cite{li2014discrete}. 

\section{Results and Discussion \label{sec:Simulation-results}}

This section presents the performance of the proposed FTSM controller
in simulation for the mirror-based pointing sensor (Fig. \ref{fig:Mirror-based-sensor})
with the identified system parameters presented in Table \ref{tab:The-identified-parameters-H11-H22}.
Two cases were considered respectively for step and sinusoidal reference signals.
In the simulation, gains of the controller were set as: $K=$10, $\alpha=1,$
$\beta=2$, $q_{1}=7$, $p_{1}=9,$ and $\Phi=70.$ Sampling period
for the controller was set to 0.01s. In addition, disturbances of
0.1 degree was also added to the system to judge the
robustness.

\subsection{Tracking Response}
\begin{figure}[h]
\begin{centering}
\vspace{-0.5 cm}
\includegraphics[width=0.9\columnwidth]{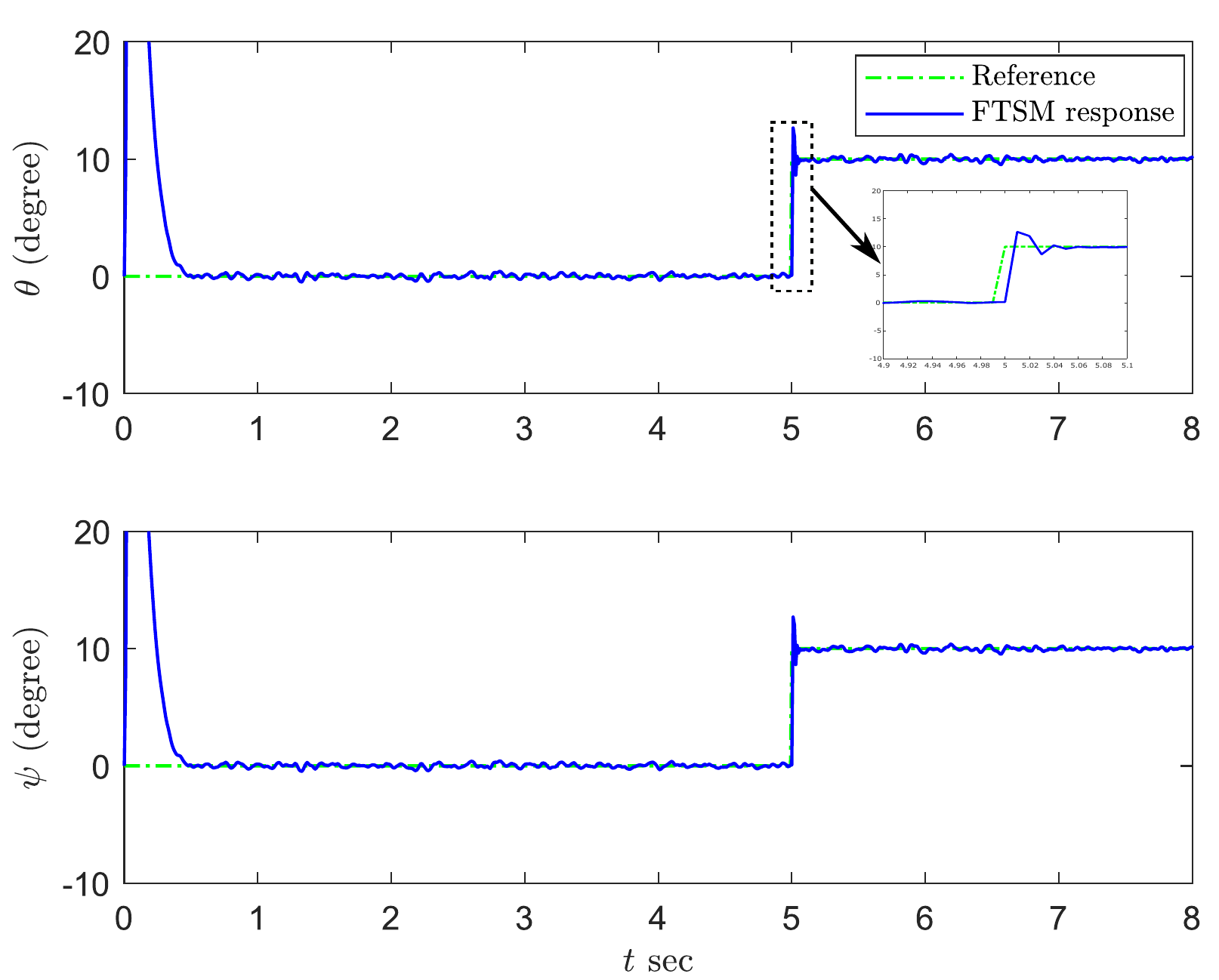}

\par\end{centering}

\caption{ Step signal tracking by the FTSM controller.\label{fig:Step-signal-tracking-FTSM}}
\end{figure}

The step value for the reference signal is 10$^{o}$. Figure \ref{fig:Step-signal-tracking-FTSM}
shows the tracking respose of controller, which clearly indicates
that the system is able to track the reference signal. Settling time
for the azimuth and elevation is around 0.04 sec, which can be observed
in the zoomed section of the step response. 

\begin{figure}[h]
\begin{centering}
\includegraphics[width=0.9\columnwidth]{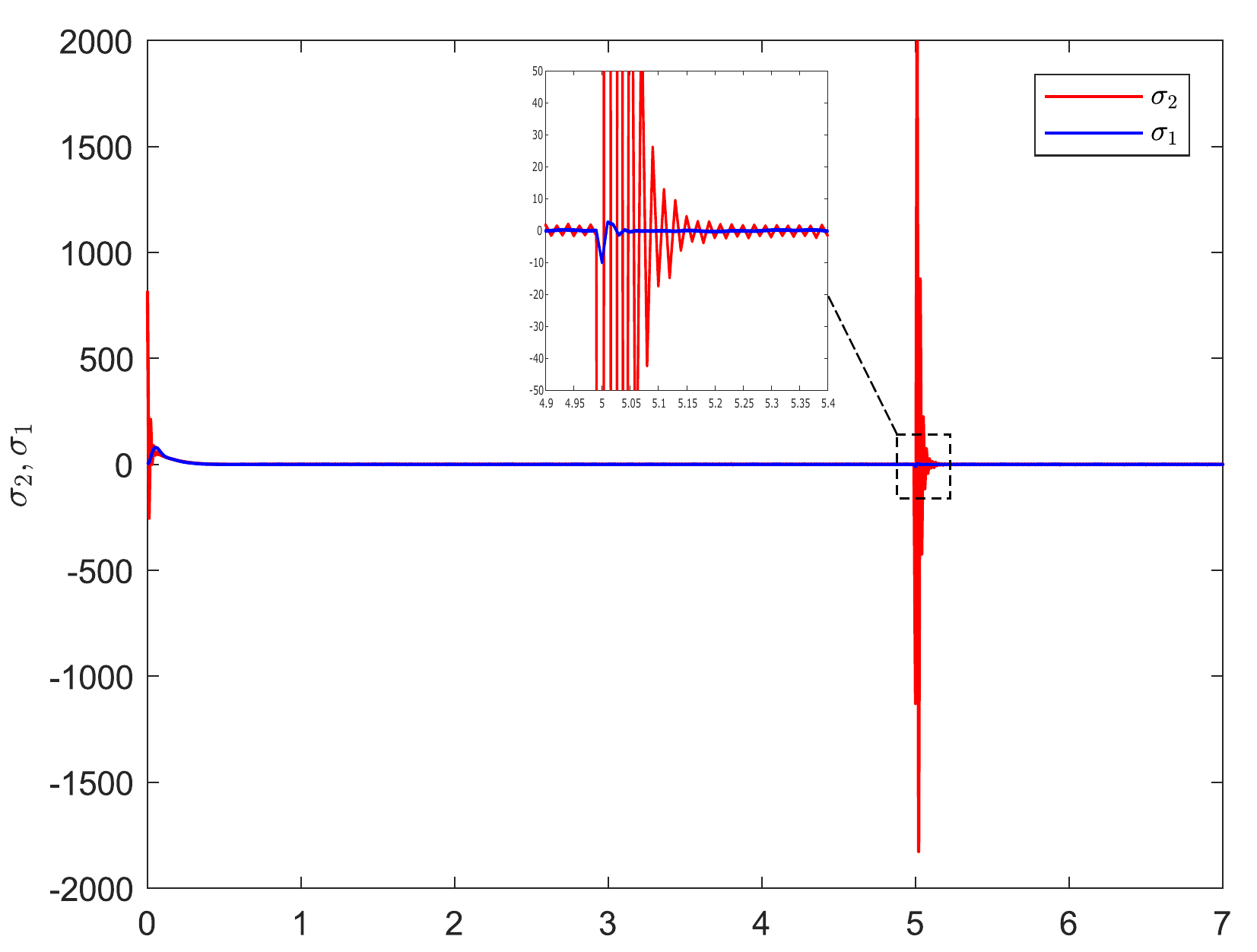}
\par\end{centering}
\caption{$\sigma_{1}$ and $\sigma_{2}$ for elevation angle.\label{fig:s1-and-s2-Plots}}
\end{figure}
Figure \ref{fig:s1-and-s2-Plots} shows the sliding functions $\sigma_{1}$
and $\sigma_{2}$ for the elevation angle. From the figure it is clear
that $\sigma_{2}$ converges towards the sliding mode, that is $\sigma_{2}=0.$
The signal, then, remains bounded within the region $-10<\sigma_{2}<10$.
Furthermore, it can also be observed that $\sigma_{1}$ is also bounded
in the steady-state condition, which is much lower than $\sigma_{2}.$ 

\begin{figure}[h]
\begin{centering}
\includegraphics[width=0.9\columnwidth]{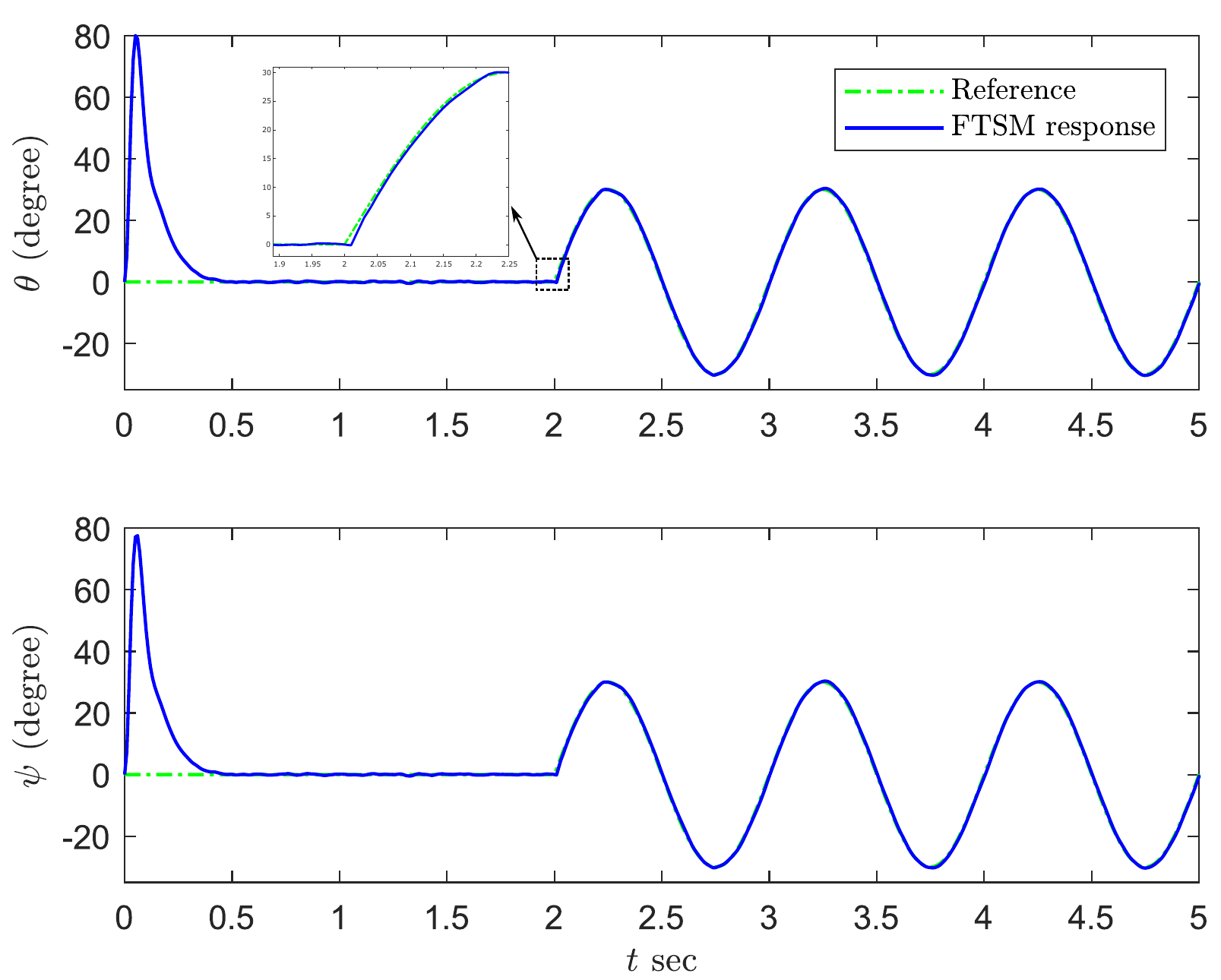}
\par\end{centering}
\caption{Sinusoidal tracking by the system.\label{fig:Sinusoidal-tracking-by-system}}
\end{figure}

Tracking response of the system for sinusoidal signal is presented
in Fig. \ref{fig:Sinusoidal-tracking-by-system}. The amplitude
and frequency of the reference are $30^{o}$ and 1Hz, respectively.
The tracking response of the system is close to the step signal which
can be verified by observing at the zoomed section of the plot.

\subsection{Comparison with Model Predictive Control and Terminal Sliding Mode}

\begin{figure}[h]
\vspace{-0.5cm}
\begin{centering}
\includegraphics[width=0.9\columnwidth]{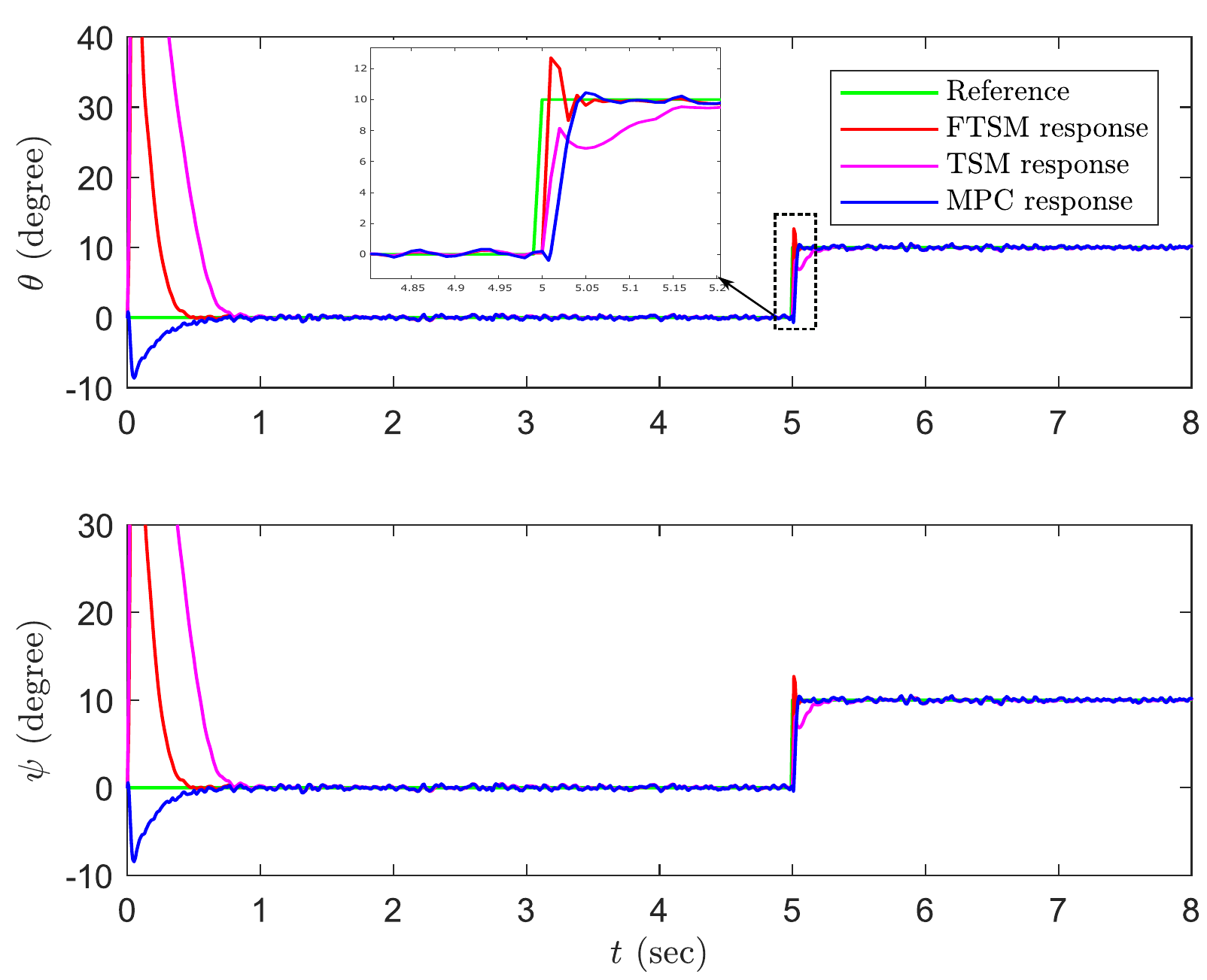}
\par\end{centering}
\caption{Comparison of the proposed method with TSM and MPC for step
signal\label{fig:Comparison-of-the-methods-Step}}
\end{figure}
\begin{figure}[H]
\begin{centering}
\includegraphics[width=0.9\columnwidth]{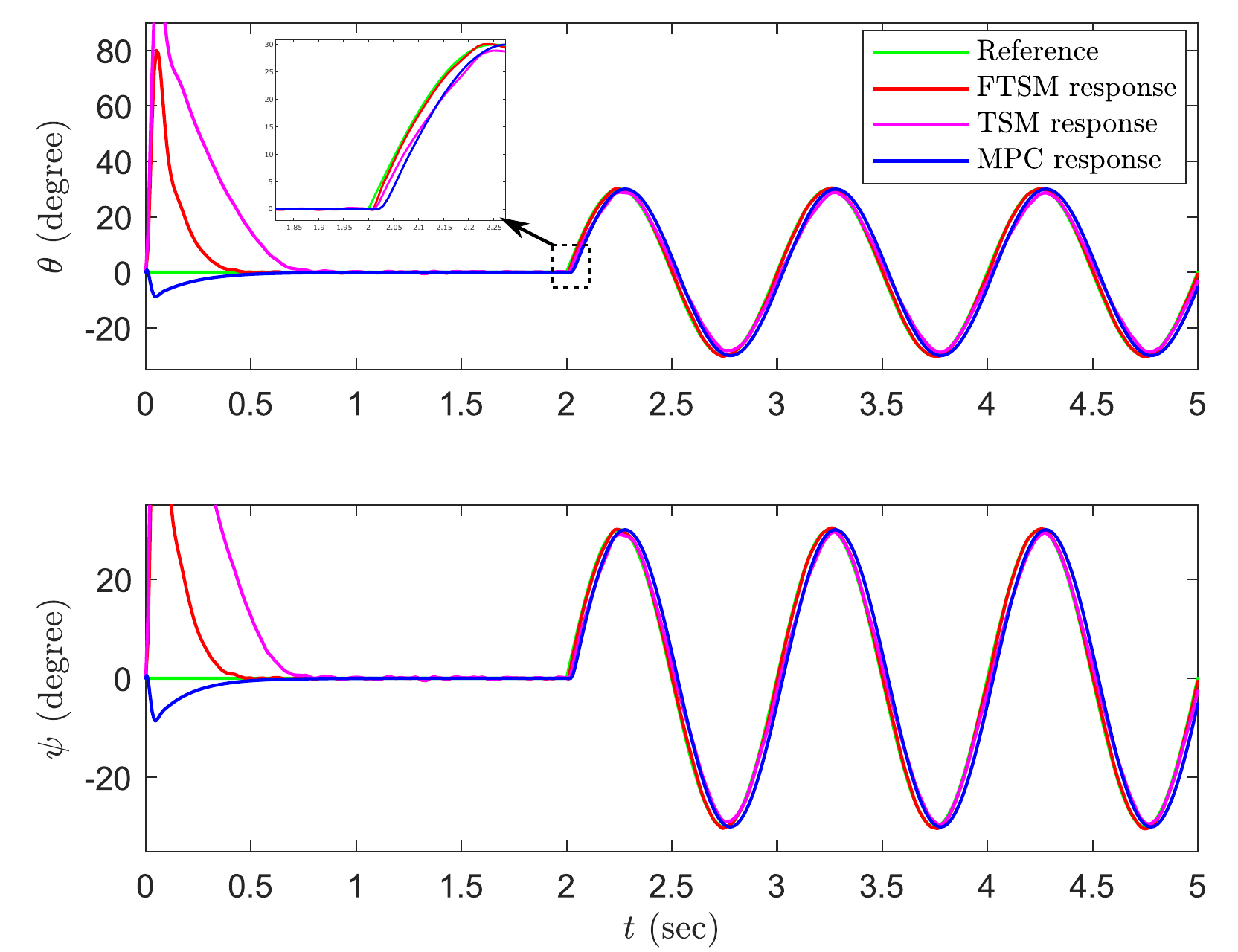}
\par\end{centering}
\caption{Comparison of the proposed method with TSM and MPC for sinusoidal
signal\label{fig:Comparison-of-the-methods-sin-ref}}
\end{figure}
\begin{figure}[H]
\begin{centering}
\vspace{-0.1 cm}
\includegraphics[width=0.9\columnwidth]{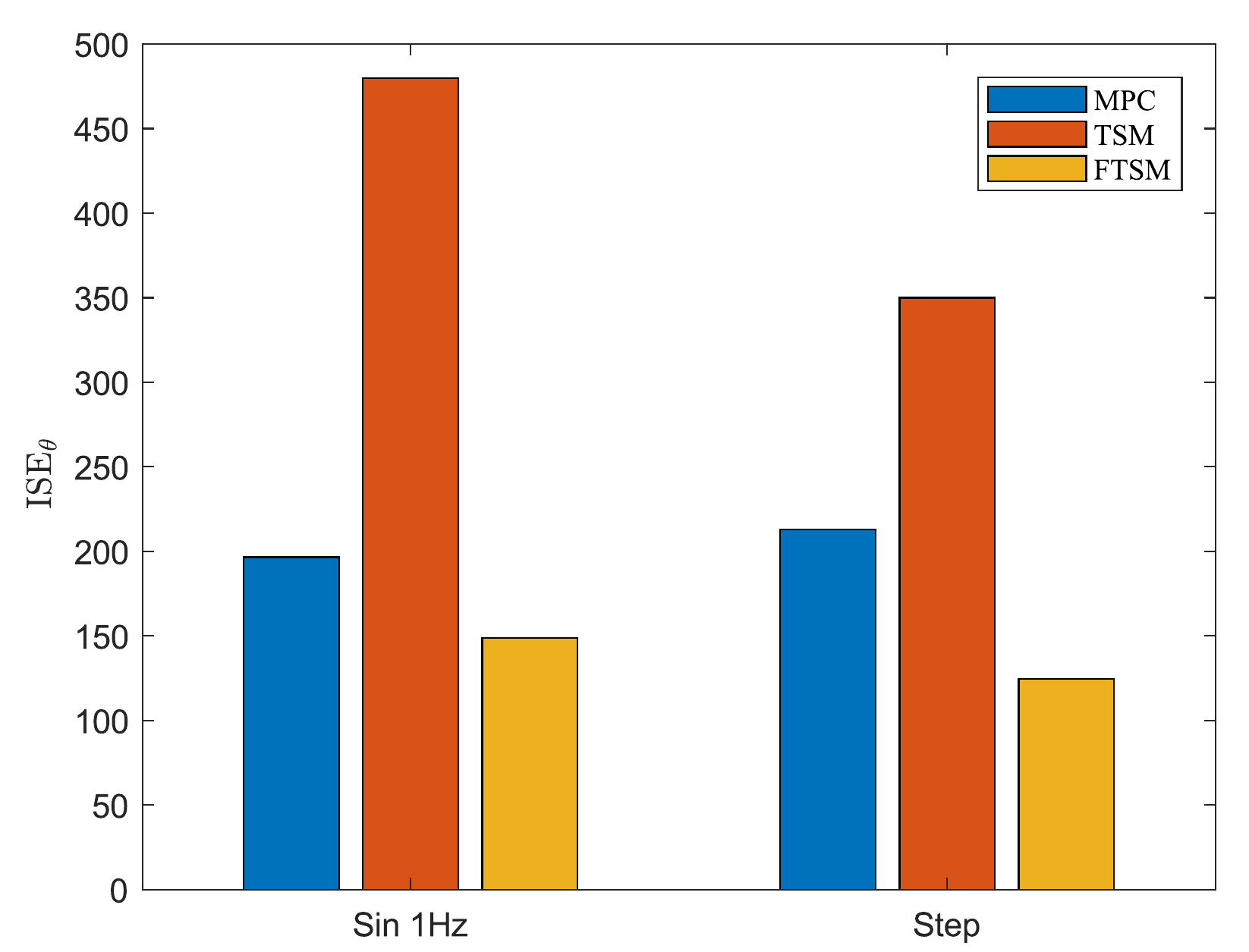}
\par\end{centering}
\caption{Comparison of ISE for elevation angle\label{fig:Comparison-of-ISE-theta} }
\end{figure}
\begin{figure}[H]
\begin{centering}
\includegraphics[width=0.95\columnwidth]{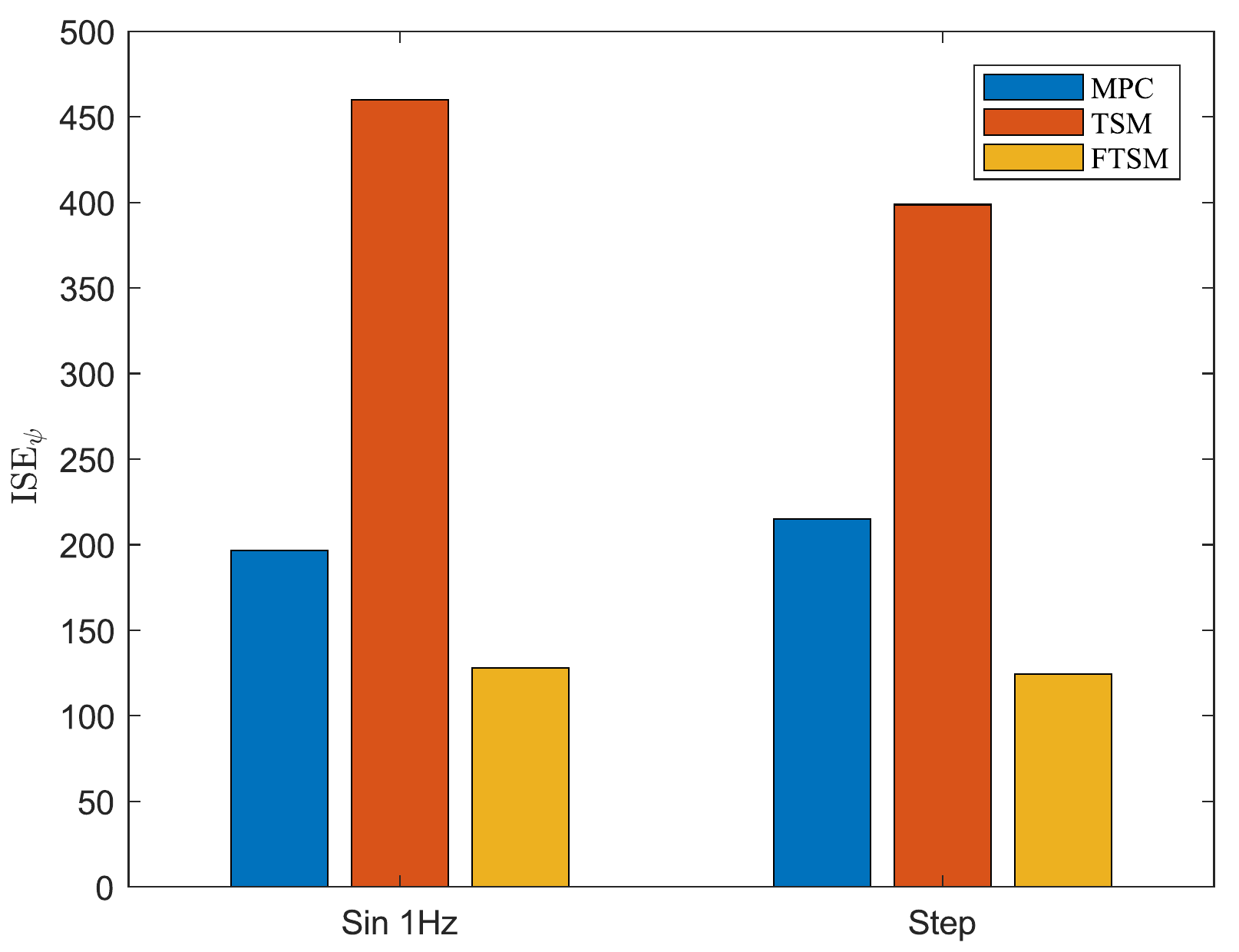}
\par\end{centering}
\caption{ISE for azimuth angle.\label{fig:Comparison-of-ISE-azimuth}}
\end{figure}

Figures \ref{fig:Comparison-of-the-methods-Step} and \ref{fig:Comparison-of-the-methods-sin-ref}
represent the comparison of our method with Terminal Sliding Mode (TSM)
and Model Predictive Control (MPC) for step and sinusoidal reference
signals.
The details on the design and architecture of the MPC for
the mirror based pointing system can be found in \cite{Singh2018}. Similarly, for TSM, we implemented the control law presented in \cite{li2014discrete}.
The tracking responses of the system clearly indicates that the proposed
control law is faster than the other methods which can be observed
in the zoomed section of the plots. For instance, the settling time for
the proposed control law while tracking the step signal is 0.04 sec, compared
to the 0.5 sec and 0.05 sec, respectively, for TSM and MPC. Furthermore,
the response of the FTSM for the sinusoidal signal is close to the reference
signal compared to other methods. 
In order to evaluate the steady state error for all the control methods
we consider the Integral of Square Error (ISE) performance index. The
ISE performance index is defined as
\begin{equation}
\text{ISE}=\sum_{n=1}^{M}e^{2}[n],
\end{equation}
where $e[n]$ is the difference between actual and reference signal.
Plots for the ISE for both elevation and azimuth angles are presented
in Fig. \ref{fig:Comparison-of-ISE-theta} and \ref{fig:Comparison-of-ISE-azimuth},
respectively. The ISE clearly indicates that the FTSM has lowest values
for both scenarios, i.e step and sinusoidal signals. 

\subsection{Validation and Verification with Experiment Results}

This subsection presents the comparison of the proposed control method
with the results obtained from the real-time experiment on the mirror-based
pointing sensor (Fig. \ref{fig:Mirror-based-sensor}). The results of the real-time
experiment were presented in our previous study \cite{Singh2018},
and interested readers can refer to the research article for the details
on the experimental setup. 

\begin{figure}[H]
\begin{centering}
\includegraphics[width=0.9\columnwidth]{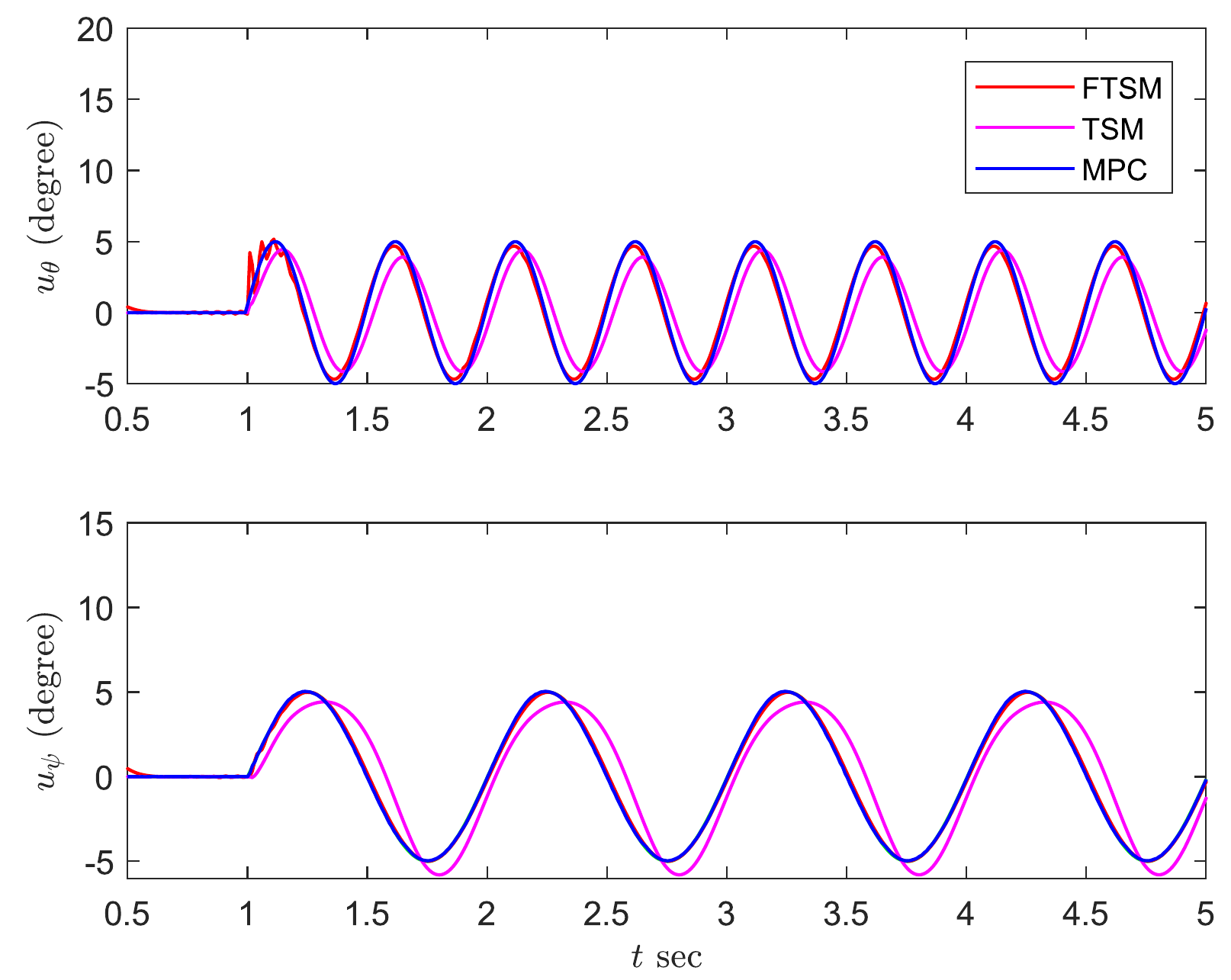}
\par\end{centering}
\caption{Comparison of control signals. \label{fig:Comparison-of-control-signals} }

\end{figure}

Figure \ref{fig:Comparison-of-control-signals} compares the control
signals generated by the FTSM and TSM with the one obtained during
the experiment. From the figure it is clear the control signals generated
by proposed method are close to the actual signal. Hence, it signifies
the effectiveness of the proposed method in real-time experiment. 

Comparison of the ISE values of the controllers are presented in Table
\ref{tab:Comparison-of-ISE}. Here also, ISE values for both elevation
and azimuth angles are low for the proposed method compared to other
methods. For instance, ISE of elevation angle for the proposed method
is 22.5, compared to 33 and 874 for MPC and TSM, respectively. 
\vspace{0.2 cm}
\begin{table}[H]

\begin{centering}
\caption{Comparison of ISE. \label{tab:Comparison-of-ISE}}
\begin{tabular}{c|c|c}
 & $\theta$ & $\psi$\tabularnewline
\hline 
FTSM & 22.5 & 2.6\tabularnewline
\hline 
MPC & 33.7 & 3.9\tabularnewline
\hline 
TSM & 874 & 833\tabularnewline
\hline 
\end{tabular}
\par\end{centering}

\end{table}

In summary, these results highlight the merits of the proposed method over TSM and MPC, in terms of ISE performance index, in all the scenarios.
\vspace{0.4 cm}

\section{Conclusion \label{sec:Conclusion}}
This paper has presented an effective discrete-time Fast Terminal Sliding Mode controller for mirror-based pointing sensor, which has been applied on the decoupled state-space models of the system. Stability of the discrete-time system could be verified in terms of a reaching condition. In addition, this paper contributes to the improvement in the sliding dynamics of the system. Effectiveness of the proposed method has been verified by extensive simulations, followed by the comparisons with MPC and TSM. The controller has also shown improved performance when compared with the data obtained from real-time experimements conducted on the system using MPC. Finally, further analysis of the control system, and the establishment
of boundary region in the steady state condition remain as a future
work.
\vspace{0.8 cm}
\section*{Acknowledgment}
Authors would like to acknowledge Ocular Robotics Pty. Ltd. for their support to gather data for this paper.
\vspace{0.9 cm}
\bibliographystyle{ieeetr}
\bibliography{ICARCV}


\end{document}